\documentclass[12pt]{article}
\usepackage{graphicx}
\begin{document}
\rightline{NKU-2012-SF1}
\bigskip
\begin{center}
{\Large\bf Schwarzschild black hole surrounded by quintessence: Null geodesics}

\end{center}
\hspace{0.4cm}
\begin{center}
Sharmanthie Fernando  \footnote{fernando@nku.edu} \\
{\small\it Department of Physics \& Geology}\\
{\small\it Northern Kentucky University}\\
{\small\it Highland Heights}\\
{\small\it Kentucky 41099}\\
{\small\it U.S.A.}\\

\end{center}

\begin{center}
{\bf Abstract}
\end{center}

\hspace{0.7cm} 

We have studied the null geodesics of the Schwarzschild black hole surrounded by  quintessence matter. Quintessence matter is a candidate for dark energy. Here, we have done a detailed analysis of the geodesics and exact solutions are presented in terms of Jacobi-elliptic integrals  for all possible energy and angular momentum of the photons. The circular orbits of the photons are studied in detail. As an application of the null geodesics, the angle of deflection of the photons are computed.

{\it Key words}: static,  black holes, geodesics, dark energy, quintessence

\section{Introduction}

It is well known that we live in a universe  undergoing a period of accelerating expansion. This indicates  the presence of mysterious form of repulsive gravity called dark energy. There are several observations that support dark energy. One is the studies of type Ia supernova \cite{perl} \cite{riess}. The other is the observations related to the cosmic microwave background(CMB) \cite{sper} and large scale structure (LSS) \cite{teg} \cite{sel}.

The nature of dark energy is yet to be understood. There are several cosmological models proposed in which the dominant component of the energy density has negative pressure. One of them is  the cosmological constant $\Lambda$  which corresponds to the case of dark energy with a state parameter $ \omega_q =-1$. However, there is a key problem that is yet to be understood about the cosmological constant from the point of view of fundamental physics. Its observed value is too small which is also called the fine-tuning problem \cite{wein}.

There are alternative models that are proposed as candidates for dark energy. Most of these models are based on a scalar field. Such scalar field models  include but not limited to, quintessence 
\cite{carroll}, chameleon fields \cite{khoury}, K-essence \cite{amer},   tachyon field \cite{pad}, phantom dark energy \cite{cald} and dilaton dark energy \cite{gas}. For a detailed review of various models of dark energy see \cite{copeland}.

Given the fact that dark energy contains about $70\%$ of the universe and black holes are also accepted as part of our universe, studies of black holes surrounded by dark energy takes an important place in research. In this paper, we study the Schwarzschild  black hole surrounded by quintessence matter. Quintessence is described as a scalar field coupled to gravity with a potential that decreases as the field increases \cite{carroll}\cite{copeland}.  Kiselev \cite{kiselev} derived black hole solutions surrounded by quintessence matter which has the state parameter in the range, $ -1 < \omega_q < - \frac{1}{3}$. In this paper, we will focus on the solution derived by Kiselev.

There are several works done on the black hole derived by Kiselev in the literature. The quasinormal modes of the black holes  has been computed extensively. The quasinormal modes of the Schwarzschild black hole surrounded by the quintessence matter has been computed in \cite{chen} \cite{zhang1} \cite{zhang2}
\cite{ma}. The quasinormal modes of the Reissner-Nordstrom black hole surrounded by the quintessence matter has been presented in \cite{nijo1}. Hawking radiation of d-dimensional  extension of the Kiselev black hole has been studied by Chen et. al. in \cite{chen2}.

The main objective of this paper is to study the geodesic structure of massless particles of the Schwarzschild black hole surrounded by quintessence matter. The motion of particles around black holes has been studied extensively in the literature for all types of black holes. Due to the large volume of papers related to this subject, we will avoid referring the papers here. Motion of particles around a black hole is one way to understand the gravitational field around a black hole. Given the fact that dark energy is one of the most important issues need to be resolved in physics, the studies of particles around a black hole immersed in dark energy take an important place.

The paper is presented as follows: In section 2 the black hole solutions surrounded by quintessence matter is  introduced. In section 3 the general formalism  of the geodesics are given. In section 4, the radial geodesics are discussed. In section 5, the geodesics with angular momentum are discussed. In section 6, the geodesics are analyzed with a new parameter. In section 7, the applications of the null geodesics are given. Finally, the conclusion is given in section 8.


\section{Schwarzschild black hole  surrounded with\\ 
quintessential matter}

In this section we will give an introduction to the 
Schwarzschild black hole solutions  surrounded by quintessential matter obtained by Kiselev\cite{kiselev}. The geometry of such a black hole has
 the metric of the form,
\begin{equation}
ds^2 = -g(r) dt^2 + g(r)^{-1} dr^2 +  r^2 ( d \theta^2 + sin^2\theta d \phi^2)
\end{equation}
where,
\begin{equation}
g(r)= 1 - \frac{ 2 M} { r}   - \frac{ c}{ r^{ 3 w_q + 1} }
\end{equation}
Here, $M$ is the mass of the black hole and $w_q$ is the  state parameter of the quintessence mater. $c$ is a normalization factor. The parameter $ w_q$ has the range,
\begin{equation}
-1 < w_q < -\frac{1}{3}
\end{equation}
The equation of state for the quintessence mater is given by,
\begin{equation}
p_q = w_q \rho_q
\end{equation}
and
\begin{equation}
\rho_q = - \frac{ c} { 2} \frac{ 3 w_q}{ r^{ 3( 1 + w_q)}}
\end{equation}
Here $p_q$ is pressure and $\rho_q$ is the energy density. As described in \cite{kiselev}, to cause acceleration, the pressure of the quintessence matter  $p_q$ has to be  neagtive. The mater energy density $\rho_q$ is positive. Hence the parameter $c >0 $  for negative $w_q$. For more details on the derivation of the solutions and the basis to choose the parameters as given, reader is referred to the paper by Kiselev \cite{kiselev}.

One can observe that for $w_q = -1$, the function $g(r)$  for the metric reduces  to 
\begin{equation}
g(r)= 1 - \frac{ 2 M} { r}   -  c r^2
\end{equation}
which is the Schwarzschild-de-Sitter black hole space-time.

In this paper, we will focus on the special case of $w_q = -\frac{2}{3}$. Then,
\begin{equation}
g(r)= 1 - \frac{ 2 M} { r}   -  c r
\end{equation}
For $ 8 Mc <1$, the metric with the above $g(r)$ given in eq.(7) has two horizons as,
 
\begin{equation}
r_{in} = \frac{ 1 - \sqrt{ 1 - 8 M c} }{ 2 c}
\end{equation}
\begin{equation}
r_{out} = \frac{ 1 + \sqrt{ 1 - 8 M c} }{ 2 c}
\end{equation}
The inner horizon $r_{in}$ is like the Schwarzschild black hole horizon. The outer horizon $r_{out}$ is a cosmological horizon similar to what is observed in the Schwarzschild-de-Sitter black hole. Notice for small $c$, $r_{out} \approx \frac{1}{c}$ which is similar to $r_{out} \approx \sqrt{ \frac{3}{\Lambda}}$ for small $\Lambda$ in the Schwarzschild-de-Sitter case. There is a static region between the two horizons. Hence, in considering the motion of photons, we will focus on the region between the two horizons.

If $ 8 M c =1$, the space-time has a degenerate solution at $r =\frac{1}{ 2c}$. For $ 8 M c >1$, there are no horizons. All  three situations are represented in  Fig. 1.

\begin{center}
\scalebox{.9}{\includegraphics{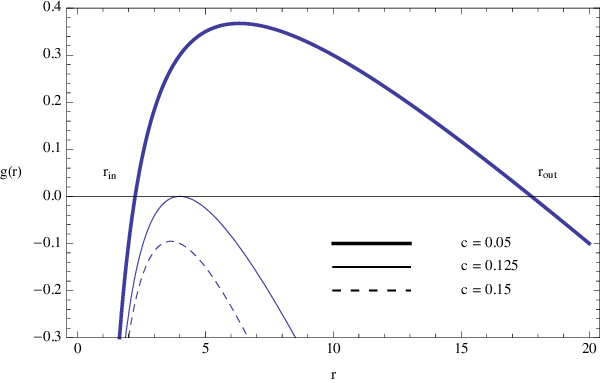}}\\
\vspace{0.1cm}
\end{center}
Figure 1. The graph shows the relation of $g(r)$ with $r$  for various values of the parameter $c$. Here, $M =1$\\

The Hawking temperature for this black hole is,
\begin{equation}
T_{r_{in},r_{out}} = \frac{1}{ 4 \pi} \left| \frac{ dg_{tt}}{dr} \right|_{r = r_{in,out}}
\end{equation}
which leads to,
\begin{equation}
T_{r_{in}, r_{out}} = \frac{ 1}{ 4 \pi} \left| \frac{ 2 M}{ r^2_{in,out}} - c \right|
\end{equation}
The temperature for both horizons are plotted in  Fig.2. The temperature of the outer horizon is smaller than the inner horizon.

\begin{center}
\scalebox{.9}{\includegraphics{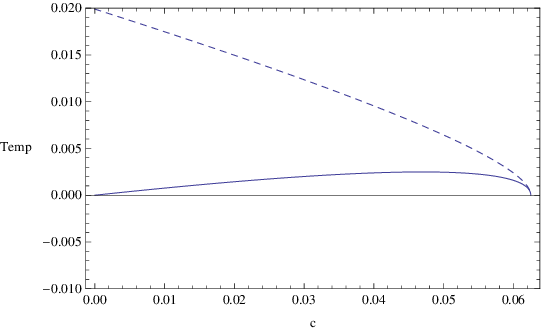}}\\
\vspace{0.1cm}
\end{center}
Figure 2. The graph shows the relation of the temperature with $c$  for both horizons. The dashed curve is for the inner horizon while the thick curve is for the outer horizon. Here, $M =2$\\


\section{ Geodesics of the black hole}

We will derive the geodesic equations for neutral particles around the quintessence  black hole. For a static spherically symmetric black hole studied in this paper, the Lagrangian ${\cal{L}}_g$ i given by,

\begin{equation}
{\cal{L}}_g =  - \frac{1}{2} \left( - g(r) \left( \frac{dt}{d s} \right)^2 +  \frac{1}{g(r)}\left( \frac{dr}{d s} \right)^2 + r^2 \left(\frac{d \theta}{d s} \right)^2 + r^2 sin \theta ^2 \left( \frac{d \phi }{d s} \right)^2 \right)
\end{equation}
There are two constants of motions (due to the fact that there are two Killing vectors $\partial_t$ and $\partial_{\phi}$) as,
\begin{equation}
g(r) \dot{t} = E_n
\end{equation}
\begin{equation}
r^2 sin^2 \theta \dot{\phi} = L
\end{equation}
We choose $\theta = \pi/2$ and $\dot{\theta} = 0$ as the initial conditions. Hence $\ddot{\theta} =0$. $\theta$ will remain at $\pi/2$ and the geodesics will be described in an invariant plane at $ \theta = \pi/2$.
With $\dot{t}$ and $\dot{\phi}$ given by equations(13) and (14), the Lagrangian in eq.(12) simplifies to be,
\begin{equation}
\dot{r}^2 + g(r) \left(  \frac{L^2}{r^2} + \epsilon   \right)  = E_n^2
\end{equation}
Here, $2 {\cal{L}}_g=\epsilon$. $\epsilon =1$ corresponds to a massive particle and $\epsilon =0$ corresponds to a massless particle. $s$ can be identified as the proper time.

One can write eq.(15) as $\dot{r}^2 + V_{e}= E_n^2$,  where,
\begin{equation}
V_{e} = \left(  \frac{L^2}{r^2} + \epsilon  \right) g(r) 
\end{equation}
By combining equations(13) and (14), one can get a relation between $\phi$ and $r$ as follows;
\begin{equation}
\frac{ d \phi} {d r} =  \frac{L}{r^2} \frac{1}{ \sqrt{  ( E_n^2 - V_{e})} }
\end{equation}
In this paper, we will only focus on null geodesics. Hence $\epsilon =0$ and the effective potential corresponds to,
\begin{equation}
V_{e} = L^2 \frac{g(r)}{r^2}
\end{equation}


\section{ Radial null geodesics}

For radial geodesics $L =0$. Hence the effective potential,
\begin{equation}
V_{e} = 0
\end{equation}
Hence, the  two equations for $\dot{t}$ and $\dot{r}$ simplifies to,
\begin{equation}
\dot{ r} = \pm E; \hspace{1 cm} \dot{t} = \frac{E_n}{g(r)}
\end{equation}
By combining above two equations one obtain,
\begin{equation}
\frac{ dt}{ dr} = \pm\frac{1}{g(r) } = \pm \frac{ 1} { (1 - \frac{ 2 M}{ r}-c r) }
\end{equation}
eq.(21) can be integrated to give the coordinate time $t$ as a function of $r$ as,
\begin{equation}
t = \pm \left( \frac{-2}{ \sqrt{ 1 - 8 cM} } Tanh^{-1}\left( \frac{ - 1 + 2 cr}{\sqrt{ 1 - 8 M c}} \right) + Log \left( r - 2 M  - c r^2 \right)\right) + const_{\pm}
\end{equation}
The constant of integration is imaginary. However, when one compute the time $t$, the overall expression will be real.  The proper time $s$ can be obtained by integrating,
\begin{equation}
\frac{ d s}{ dr} = \pm\frac{1}{ E_n} 
\end{equation}
which leads to,
\begin{equation}
 s = \pm \frac{ r}{ E_n} + const_{\pm}
\end{equation}
When $ r \rightarrow r_{in}$, $ t \rightarrow \infty$. When 
$ r \rightarrow r_{in}$, $ s \rightarrow   \frac{ r_{in}} { E} $, which is finite.  Therefore,  the proper time $s$ is finite while the coordinate time$t$ is infinite. 
Similar results were obtained for the   Schwarzschild black hole in \cite{chandra}. Fig.3 represents both $s$ and $t$  to show this. Note that we have  picked the plus sign in this case, since we are studying the ingoing light rays.

\begin{center}
\scalebox{.9}{\includegraphics{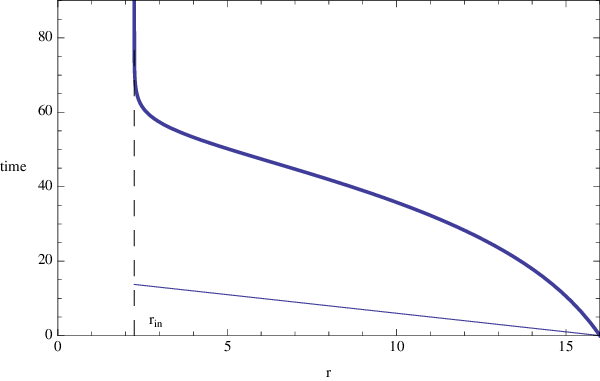}}\\
\vspace{0.1cm}
\end{center}
Figure 3. The graph shows the relation of the coordinate time $t$
and the proper time $s$  with the radius $r$. Here, $M =1, c = 0.05$. The initial position is at $r = 16$. The inner horizon is at $r_{in} = 2.254$ and the outer horizon $r_{out} = 17.746$. The dark curve is for $t$ and the light curve is for $s$.

\section{ Null geodesics with angular momentum ($ L \neq 0$)}


\subsection{ Effective potential for $L \neq 0$}

When $ L \neq 0$,
\begin{equation}
 V_{e} = g(r) \frac{L^2}{ r^2}
\end{equation}
For $ r = r_{in}$ and $r_{out}$, $ V_{e} = 0$. In  Fig. 4, the  $V_{e}$ is plotted for various values of $c$. The potential for the Schwarzschild black hole is higher than for the black hole surrounded with the  quintessence.

\begin{center}
\scalebox{.9}{\includegraphics{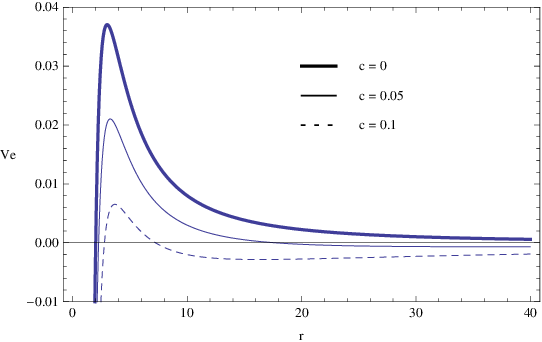}}\\
\vspace{0.1cm}
\end{center}
Figure 4. The graph shows the relation of $V_{e}$ with the parameter $c$. Here, $M =1$ and $L =1$\\

When the $V_{e}$ is expanded, it looks like,
\begin{equation}
V_{e} = \frac{L^2}{r^2} - \frac{ 2 L^2  M }{r^3} - \frac{ L^2 c}{r}
\end{equation}
The contribution to the effective potential from the quintessence is 
$-\frac{ L^2 c}{r}$ compared to the Schwarzschild black hole. Note that in the Schwarzschild-de-Sitter case, the contribution is a constant $ -L^2 \Lambda$.

Since the energy and the effective potential are related by $\dot{r}^2 + V_{e}= E_n^2$,  the motion of the particles depend on the energy levels. In  Fig.5, the effective potential is plotted and three energy levels, $E_1$, $E_c$ and $E_2$.\\
\noindent

{\bf Case 1:}  { $E_n = E_c$}

Here, $ E_n^2 - V_{e} = 0$ and $\dot{r} =0$. The orbits are   circular. Due to the nature of the potential at $ r=r_c$, these are unstable.\\
\noindent

{\bf Case 2:}  { $E_n = E_2$}

Here, $ E_n^2 - V_{e} \geq 0$ in two regions as is clear from  Fig.5. If the photons start the motion at $r > r_c$, it will fall to a minimum radius and fly back to large values of $r$. If the photons start the motion at $r < r_c$, then the photons will fall into the singularity.\\

{\bf Case 3:}  { $E_n = E_1$}

Here, $ E_n^2 - V_{e} > 0$ and $\dot{r} >0$  and  the photons coming from large $r$ values will  cross the horizon at $ r = r_{in}$ and will fall into the singularity.

\begin{center}
\scalebox{.9}{\includegraphics{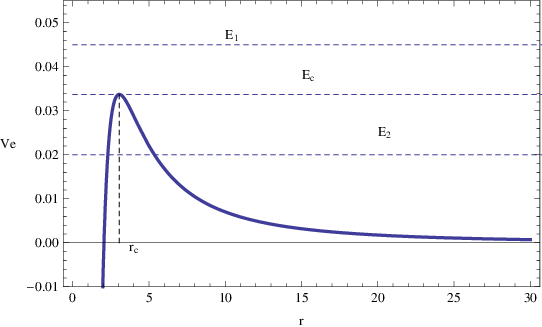}}\\
\vspace{0.2cm}
\end{center}
Figure 5. The graph shows the relation of $V_{e}$ with the energy $E_n$. Here, $ M =1, c = 0.01$ and $L =1$.

\subsection{ Circular orbits}

Circular orbits ocuure when $E_n=E_c$ as explained in section(5.1). Hence at $r=r_c$ and $\dot{r} =0$. Hence
\begin{equation}
V_{e} = E_c^2
\end{equation}
and
\begin{equation}
\frac{ d V_{e}}{dr} =0
\end{equation}
From eq.(28),  solutions for circular orbit radius $r$ are obtained as,
\begin{equation}
r_{\pm} = \frac{1}{c} \left( 1  \pm \sqrt{ 1 - 6 c M} \right)
\end{equation}
The larger root $r_+$ is greater than $r_{out}$ and $ r_{in} <  r_{-} < r_{out}$. Since the motion of interest is between the horizons, the radius of circular orbits considered will be $r_-$. We will name it as $r_c$ in  the rest of the paper. Hence,
\begin{equation}
r_c  = \frac{1}{c} \left( 1  - \sqrt{ 1 - 6 c M} \right)
\end{equation}
In the Fig.6, the radius of circular orbits are given as a function of $c$. One can observe that, the radius is bigger for non-zero values of $c$.

\begin{center}
\scalebox{.9}{\includegraphics{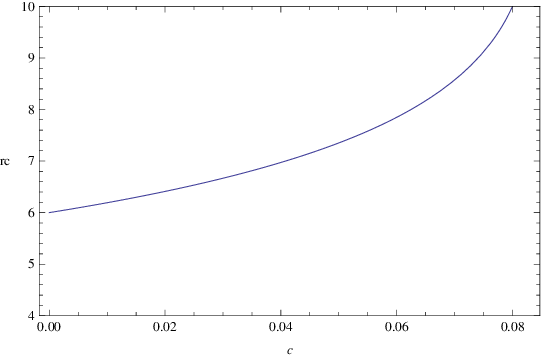}}\\

\vspace{0.2cm}
\end{center}
Figure 6. The graph shows the relation of $r_c$ with $c$. Here $M=2$.\\

Due to the nature of the potential at $r=r_c$, the circular orbits at $r=r_c$ are unstable. The hypersurface at $ r = r_c$ is known as the ``photon sphere'' \cite{vir}. When $c \rightarrow 0$, $ r_c \rightarrow 3 M$. This is the radius of the unstable circular orbit of the Schwarzschild black hole\cite{chandra}. Circular orbits takes a special place in the studies of geodesics. The null circular orbit is the boundary between two qualitatively different regions. An interesting paper on null circular orbits of a hairy black hole is given by Hod\cite{hod2}.

The energy  $E_n$ and $L$ at the circular orbit are related as,
\begin{equation}
\frac{ E_c^2}{ L_c^2} = \frac{g(r_c)} { r_c^2} = \frac{ (r_c - 2 M - c r_c^2)} { r_c^3} = \frac{ 1}{ D_c^2}
\end{equation}
Here, $D_c$ is the impact parameter at the critical stage.


\subsubsection{The Time Period}

The time period in proper time ($s$) for circular orbits can be computed from eq.(14) as
\begin{equation}
T_{s} = \frac { 2 \pi r_c^2 } { L}
\end{equation}
The values $T_s$  for the Schwarzschild black hole is given by,
\begin{equation}
T_{s, Schwarzschild} = \frac{ 2 \pi ( 3M)^2} { L}
\end{equation}
Since the radius $r_c$ is smaller for the Schwarzschild black hole,  $T_s$ is smaller for the Schwarzschild black hole.

From combining the two equations eq.(13) and eq.(31), the time period for the coordinate time $t$ is computed as,
\begin{equation}
T_t = \frac{ 2 \pi r_c }{ \sqrt{ g(r_c)}} = \frac{ 2 \pi  r_c^{3/2}}
 { \sqrt{ r_c  - 2 M -cr_c^2}}
\end{equation}
The value for  $T_t$  for the Schwarzschild black hole is given by,
\begin{equation}
T_{ t, Schwarzschild} = 3 \sqrt{3} M
\end{equation}
The graphs for the time periods are given in Fig.7 and Fig.8. By observing the graphs of the time periods, it is clear that the periods for the Schwarzschild black hole is smaller in comparison with the quintessence black hole. 
In a recent paper, Hod made an interesting observation that the null circular geodesics provide the shortest possible orbital period to circle a black hole \cite{hod1}.

\begin{center}
\scalebox{.9}{\includegraphics{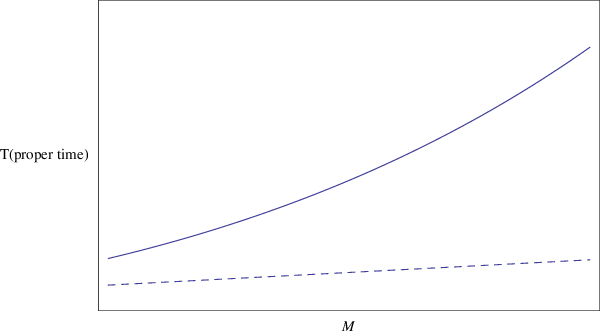}}\\

\vspace{0.2cm}
\end{center}
Figure 7. The graph shows the relation of $T_{s}$ with the mass $M$. The dark curve is for $T_{s, quintessence}$ and the dashed curve is for $T_{s,Sch}$. Here, $c =0.05$ and $L =1$.\\

\begin{center}
\scalebox{.9}{\includegraphics{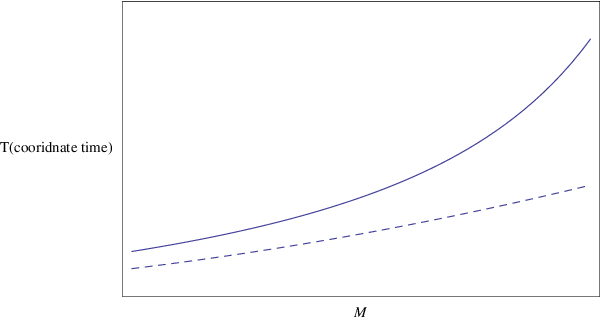}}\\

\vspace{0.2cm}
\end{center}
Figure 8. The graph shows the relation of $T_t$ with $M$. The dark curve is for $T_{t, quintessence}$ and the dashed curve is for $T_{t ,Sch}$. Here $c=0.01$ and $L=1$.\\

\subsubsection{ Lyapunov exponent for the unstable circular orbits}

The instability time-scale of the  unstable circular null geodesics is given by the Lyapunov exponent $\lambda$. The expression for $\lambda$ was derived by Cardoso et.al \cite{car} as,

$$
\lambda = \sqrt{ \frac{ -V_{e}''(r_c)}{ 2 \dot{t}(r_c)^2}} 
= \sqrt{ \frac{-V_{e}''(r_c) r_c^2 g(r_c)}{2 L^2} }
$$
\begin{equation}
= \sqrt{-c^2 - \frac{ 24 M^2}{r^4} + \frac{ 18 M}{r^3} - \frac{ 3}{r^2} - \frac{ 14 c M}{r^2} + \frac{ 4 c}{r}}
\end{equation}

\begin{center}
\scalebox{.9}{\includegraphics{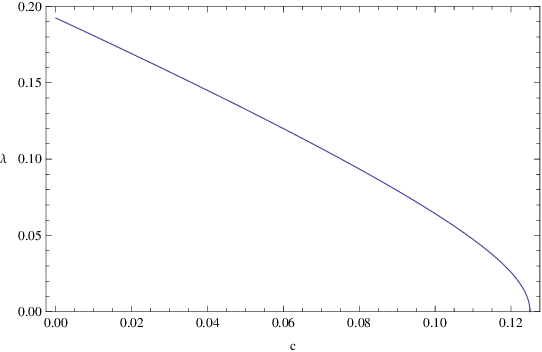}}\\
\vspace{0.2cm}
\end{center}
Figure 9. The graph shows the Lyapunov exponent $\lambda$ as a function of $c$.
Here, $M = 1$.\\

The instability of the circular orbits are more for the Schwarzschild black hole in comparison with the black hole with a non-zero $c$. In the paper by Cardoso et.al \cite{car}, a critical exponent for instability of orbits were defined as,
\begin{equation}
\gamma = \frac{ T_{\lambda}} { T_{ s}}
\end{equation}
Here, $T_{\lambda}$ is the instability time scale  which is related to the Lyapunov exponent as $ T_{\lambda} = 1/ \lambda$. Hence for large $c$ values, $\lambda$ becomes smaller for the same critical exponent. 


\subsection{Force on the Photons}

Since we have already computed the effective potential, one can obtain the effective force on the photon as,
\begin{equation}
F = -\frac{1}{2} \frac{ d V_{e}}{dr} = -\frac{ 3 M L^2}{r^4} + \frac{ L^2}{r^3} - \frac{ c L^2}{ 2 r^2}
\end{equation}
Here, we have divided $\frac{dV_{e}}{dr}$ by $2$ since the equation of motion is written as $\dot{r}^2 + V_{e}= E_n^2$.  The first term, which is the Newtonian term, is attractive since it is negative. The second term is repulsive. The third term which is the force due to the quintessence matter, is attractive. It is interesting to notice that even though dark energy in cosmology is associated with a repulsive force to facilitate acceleration, in this situation, the dark energy term is attractive.  This is a result due to the fact that we are studying a static configuration  between the two horizons. 

\begin{center}
\scalebox{.9}{\includegraphics{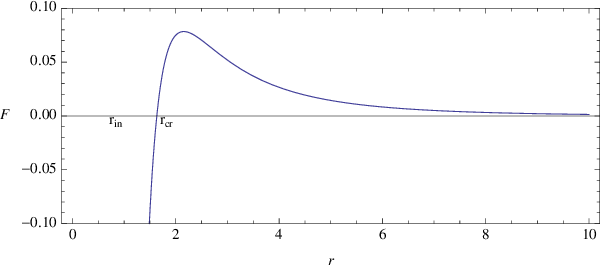}}\\

\vspace{0.2cm}
\end{center}

\begin{center}
\scalebox{.9}{\includegraphics{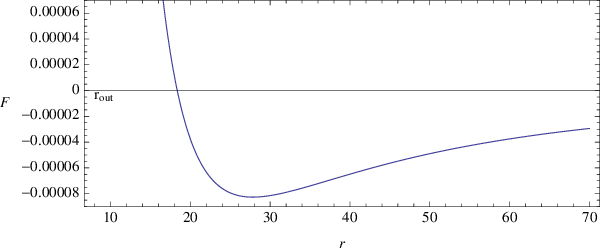}}\\

\vspace{0.2cm}
\end{center}

Figure 10. The graphs shows the relation of $F$ with $r$. Here $M=0.5, c = 0.1$ and $L=2$.\\

From  Fig.10, the force $F =0$ at,
\begin{equation}
r_{\pm} = \frac{1}{c} \left( 1  \pm \sqrt{ 1 - 6 c M} \right)
\end{equation}
One can see from the graph that $r_{out} < \frac{1}{c} \left( 1  + \sqrt{ 1 - 6 c M}\right)$. Since we are only concern about the motion for $r < r_{out}$, one can conclude that the  force is positive for 
$r_{c} < r < r_{out}$ which leads to a repulsive force. On the other hand, for $ r < r_c$, the force is negative leading to an attractive force.

The maximum repulsive force occur at $ r = r_{max}$ where,
\begin{equation}
r_{max} = \frac{ 3}{ 2 c}  \left( 1 - \sqrt{ 1 - \frac{16 c M }{3}} \right)
\end{equation}


\section{Analysis of the geodesics with the variable $ u = \frac{1}{r}$}

Geodesics equation of motion can also be studied  using a well known change of variable $ u = \frac{1}{r}$. Then the eq.(17) becomes,
\begin{equation}
\left(\frac{ d u}{ d \phi} \right)^2 = \Psi(u)
\end{equation}
where,
\begin{equation}
\Psi(u) =  2 M u^3  -   u^2  + c u  + \frac{ E_n^2}{L^2}
\end{equation}
The geometry of the geodesics depends on the  roots of the equation $ \Psi(u) =0$. For any value of the parameters in the theory $ M, c, L, E_n$, the function $\Psi(u) \rightarrow - \infty$ for 
$ u \rightarrow  -\infty$ and $\Psi(u) \rightarrow + \infty$ for 
$ u \rightarrow  + \infty$.  When $ u =0$, $\Psi(u) = + \frac{E_n^2}{L^2}$.  Hence, $\Psi(u)$ definitely has a real root ($u_1$) which is negative. Let the names of 
the roots of $\Psi(u)$ be $u_1, u_2$ and $u_3$. The sum and the products of the roots $u_1, u_2$ and $u_3$ of the polynomial $\Psi(u)$ have the following relations to the coefficiants of $\Psi(u)$ as, \cite{math},
\begin{equation}
u_1 + u_2 + u_3 = \frac{1}{ 2 M}
\end{equation}
\begin{equation}
u_1 u_2 u_3 = - \frac{ E_n^2}{ 2 M L^2}
\end{equation}
Since $u_1$ is real and negative, the roots $u_2, u_3$ has to be positive if they are real. This  conclusion is based on observing the signs of eq.(43) and eq.(44). The polynomial $\Psi(u)$ has a negative real root ($u_1$) always. Therefore the  two roots ($ u_2, u_3$) will be either real or complex-conjugate.  $u_2$ and $u_3$ has to be positive if they are real. They also can be degenerate. Also, if they are real, they could be degenerate roots as well.

The function $\Psi(u)$ for general values of $ M, c, E_n$ and $L$ is given in  Fig.11. 

\begin{center}
\scalebox{0.9}{\includegraphics{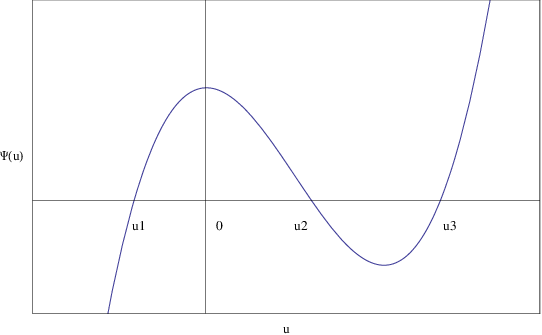}}\\
\vspace{0.2cm}
\end{center}
Figure 11. The graph shows the function $\Psi(u)$ for $ M =0.5, c = 0.01, L = 30$ and $ E_n = 9$.\\


\subsection{ General solution for the geodesics in terms of $u$}
From eq.(40) derived above,

\begin{equation}
\left( \frac{ du}{d \phi} \right) = \pm \sqrt{ \Psi(u) }
\end{equation}
where $f(u)$ can be written as,
\begin{equation}
\Psi(u) = 2 M ( u - u_1) ( u - u_2) ( u - u_3)
\end{equation}
The ``+'' sign is chosen  without lose of generality. The equation $ \frac{ du} { \sqrt{ \Psi(u) }} = d \phi$ is integrated to obtain the  relation
\begin{equation}
\phi = \frac { - 2   \Gamma ( \mu, z)}{ \sqrt{ 2 m ( u_2 - u_1)}} + constant
\end{equation}
Here,
\begin{equation}
sin \mu  = \sqrt{ \frac{ ( u_2 - u_1)}{ ( u  - u_1) }}
\end{equation}
\begin{equation}
z = \frac{ ( u_3 - u_1) }{ (u_2 - u_1)}
\end{equation}
Here, $\Gamma(\mu,z)$ is the Jacobi-elliptic integral.

Depending on the values of the $E_n, L, c, M$, the integration constant would be real or imaginary. Also, $u_1$ is always real while the nature of $u_2$ and $u_3$ depend on the values of $E_n, L, c, M$.

\subsection{ Circular orbits}

We will analyze in more detail the circular orbits here. In this case,  the function $\Psi(u)$ has a degenerate root at $ u = u_c$. Hence, $\Psi(u)$ can be written as,
\begin{equation}
\Psi(u) =  2 M ( u - u_c)^2  ( u - u_1)
\end{equation}
Here, $u_c$ is equal to $ \frac{1}{r_c}$ and  $r_c$ is given by  eq.(30). From eq.(42),
\begin{equation}
u_1 = \frac{1}{2M} - 2 u_c 
\end{equation}
A null geodesic arriving from $r=r_s > r_c$ ( or $ u = u_s < u_c$) undergo unstable circular orbits at $ r = r_c$ ( or $u = u_c$). The form of the function $\Psi(u)$ in this case is given in Fig.11.

\begin{center}
\scalebox{.9}{\includegraphics{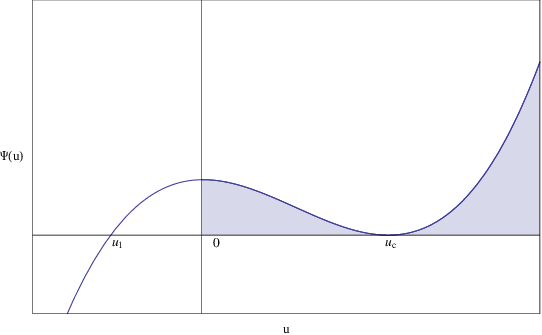}}\\

\vspace{0.4cm}
\end{center}

Figure 12. The function $\Psi(u)$ when it has degenerate roots leading to circular orbits. The shaded area represents the possible regions of motion.  Here, $M=0.5, c=0.01$ and $r_c=1.51142$.\\

When the equation $ \frac{ du} { \sqrt{ \Psi(u) }} = d \phi$ is integrated, one obtain,
\begin{equation}
u = u_1 + ( u_c - u_1)  tanh^2 \left( \frac{ \phi - \phi_0}{ a_0} \right) 
\end{equation}
$\phi_0$ and $a_0$   are  given by,
\begin{equation}
\phi_0 = a_0 tanh^{-1} \left( \sqrt{  \frac{u_1 - u_s}{u_1 - u_c} }  \right)
\end{equation}
\begin{equation}
a_0 = -\sqrt{ \frac{ 2}{ M  ( u_c - u_1) }}
\end{equation}

In  Fig.13, the polar plot of the null geodesics are given for photons arriving from $ r_s > r_c$ and having an unstable circular orbit at $ r = r_c$.

\begin{center}
\scalebox{.9}{\includegraphics{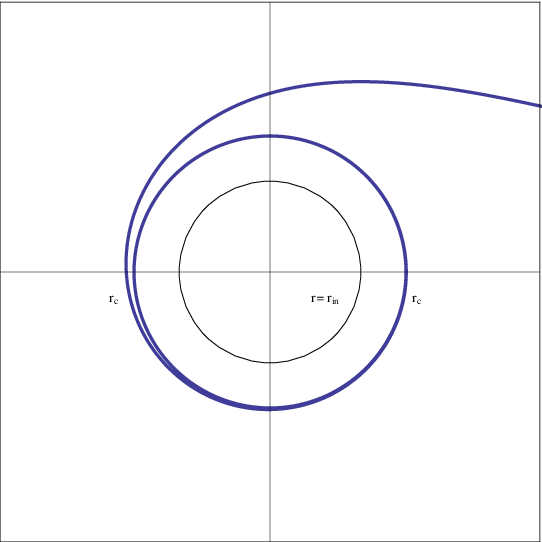}}\\

\vspace{0.4cm}
\end{center}

Figure 13. The polar plot shows the critical null geodesics approaching the black hole from $ r_s > r_c$. The geodesics have an unstable circular orbit at $r=r_c$. Here, $M=0.5, c=0.01$ and $r_c=1.51142$.\\


\subsection{ Unbounded orbits}

${\bf Case\hspace{0.1cm} 2: E_n = E_2}$

In this case, the polynomial $\Psi(u) =0$ has three real roots. Hence, photons can be in motion only in the two regions  1 and 2 in  Fig 14.

\begin{center}
\scalebox{0.9}{\includegraphics{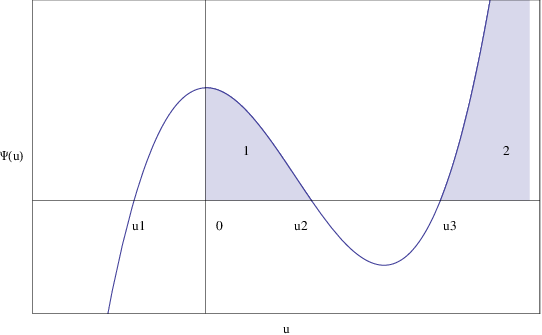}}\\

\vspace{0.2cm}
\end{center}
Figure 14. The graph shows the function $\Psi(u)$ when it has three real roots. Here $M = 0.5, c = 0.01, L = 30$ and $E_n = 9$\\

If the particle starts far from the black hole at $ r = r_s$, it will fall until $ r = r_2 =  1/ u_2$ and fly away from the black hole. The equation of motion is the same as given in eq.(46). The integration constant is chosen such that $\phi=0$ for $ r = r_s$. The corresponding motion is given in Fig 15.

\begin{center}
\scalebox{.9}{\includegraphics{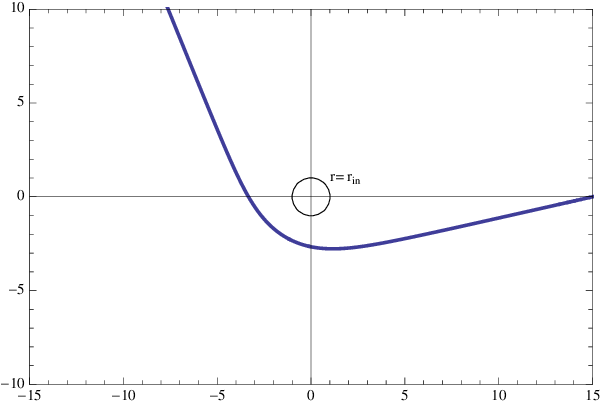}}\\

\vspace{0.2cm}
\end{center}
Figure 15. The polar plot shows the null geodesics approaching the black hole from $ r = r_s$. Here, $ M =1, c = 0.01, L = 30, E_n = 9, r_s = 15$ and  $r_2 = 2.54082$. The inner horizon is at $ r_{in} = 1.01021$\\

The second possibility  for $E_n = E_2$ corresponds to the motion starting from $r=r_3$ (or $ u = u_3$ ). Here, $r_3 < r_c$. Hence the particle will fall into the singularity crossing the horizon at $ r = r_{in}$. In this case, the solutions for  $\phi$ is given in  eq.(47).  The integration constant is computed in order for $\phi=0$ when $u = u_3$. The related motion is given in Fig.16.

\begin{center}
\scalebox{.9}{\includegraphics{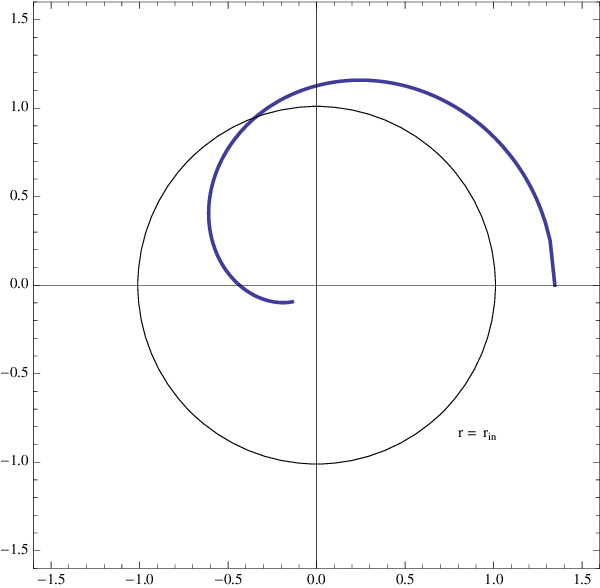}}\\
\vspace{0.2cm}
\end{center}
Figure 16. The polar plot shows the null geodesics falling into  the black hole from $r = r_3$. Here, $ M=1, c=0.01, L=30, E_n=11, r_3 = 1.34587$ and $r_{in} = 1.01021$. \\

${\bf Case \hspace{0.1cm} 3: E_n = E_1}$

Here, we will study the Case 3 given in the section(5.1). In this case, $\Psi(u) =0$ has only one real root as given in Fig 17. 
 
\begin{center}
\scalebox{.9}{\includegraphics{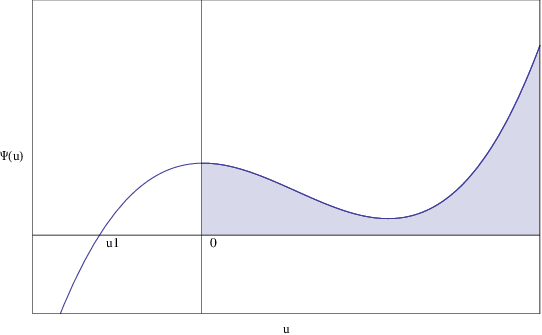}}\\

\vspace{0.2cm}
\end{center}

Figure 17. The graph shows the function $\Psi(u)$ when it has only one real root.
Here, $ M=0.5, c=0.01, L=21, E_n=9$ and  $ u_1 = -0.363391$. \\

The motion is possible in the shaded area. For all values of $r$, the photons will fall into the black hole. The equation of motion is same as in eq.(47). However, in this case, $u_2$ and $u_3$ are both imaginary and $u_1$ is real. The corresponding motion is given in Fig 18.

\begin{center}
\scalebox{.9}{\includegraphics{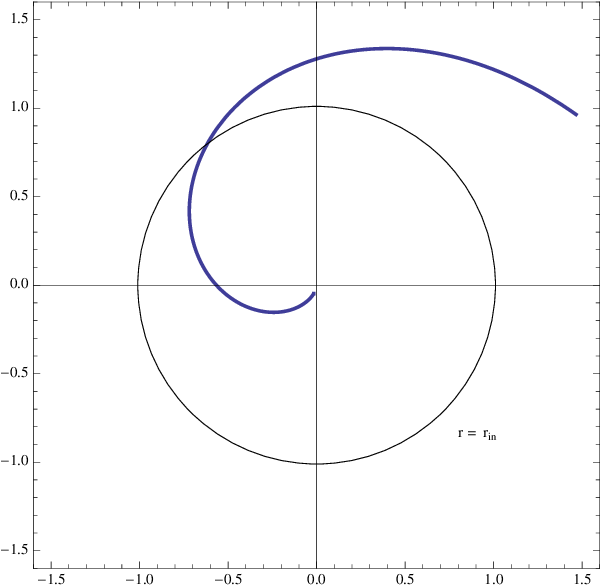}}\\

\vspace{0.2cm}
\end{center}
Figure 18. The polar plot shows the null geodesics approaching the black hole from $r = 15$ . Here, $ M=0.5, c=0.01, L=21$ and $E_n=9$. \\


\section{ Application of null geodesics: Gravitational lensing by the quintessence black hole}

Bending of light and gravitational lensing is an important aspect one can study related to null geodesics around black holes. The unbounded orbit with energy $ E_n = E_2$ studied in section(6.3) clearly represents how photons deviate from its original path when moving around a black hole. The motion is given in Fig.15.\\

\subsection{Closest approach $r_o$}

To study the bending of light, let us first calculate  the closest approach distance $r_o$ for the photon with energy $E_2$ which is defined by the value of $r$ when $ \frac{dr}{d \phi} =0$. From eq.(41) and eq.(42),
\begin{equation}
\left(\frac{1}{r^2} \frac{dr}{d \phi} \right)^2 = \Psi(r) = \frac{ 2 M}{r^3} - \frac{1}{r^2} + \frac{c}{r} + \frac{ E_n^2}{L^2}
\end{equation}
$\frac{dr}{d \phi} =0$ corresponds to the roots of $\Psi(r)=0$ which also could be written as,
\begin{equation}
r^3  + D^2 c  r^2 - D^2 r + 2 M D^2 =0
\end{equation}
The roots of the above equation corresponds to $ r_1, r_2, r_3$.  Since $r_1 <0$, we only have to consider  $r_2$ and $r_3$. From Fig.5, $r_2$ and $r_3$ are the positions where $E_2 = V_{e}$.  Clearly, $ r_3 < r_2$ (since $u_3 > u_2$). Also, $ r_3 < r_c < r_2$. Hence the closest approach is chosen to be $r_2$. 

Since eq.(56) is a cubic equation, the root $ r_2 = r_o$ can be written as,
\begin{equation}
r_o^{quintessence} = 2 \sqrt{- \frac{ \rho}{3} } cos \left( \frac{1}{3} cos^{-1} \left( \frac{ 3 \zeta}{2 \rho} \sqrt{ -\frac{ 3}{\rho}} \right) \right) - \frac{D^2 c}{3}
\end{equation}
Here, $\rho$ and $\zeta$ are given by,
\begin{equation}
 \rho = - \frac{ (D^4 c^2 + 3 D^2)}{3}
\end{equation}
\begin{equation}
\zeta = \frac{ 54 M D^2 + 9 c D^4 + 2 D^6 c^3 }{ 27}
\end{equation}
When $ c \rightarrow 0$, $r_o$ approaches the  closest approach  for the Schwarzschild black hole given by \cite{tolu},
\begin{equation}
r_o^{Schwarzschild} =  \frac{ 2 D}{ \sqrt{3}}  cos \left( \frac{1}{3} cos^{-1} \left( \frac{ -\sqrt{27} M}{D} \right) \right)
\end{equation}
Fig.19 shows the closest approach for the quintessence black hole and the Schwarzschild black hole.

\newpage

\begin{center}
\scalebox{.9}{\includegraphics{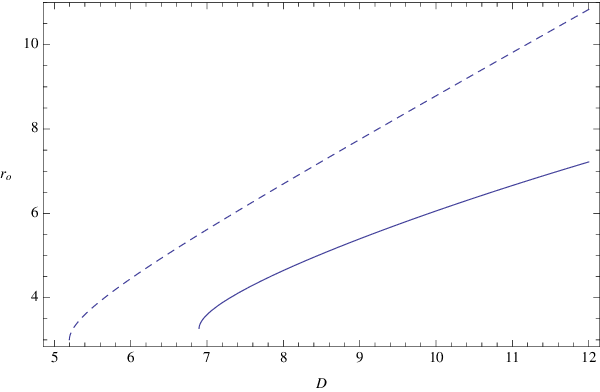}}\\
\vspace{0.2cm}
\end{center}
Figure 19. The graph shows the closest approach $r_o$ for the quintessence black hole(dark curve)  and the Schwarzschild black hole(dashed curve)  as a function of the impact parameter $D$. Here, $M = 1$  and $c = 0.05$. $D_c = 6.8956$ for the quintessence black hole and $D_c = 5.19615$ for the Schwarzschild black hole.\\


\subsection{ Bending angle}

Now, we will compute the  angle of deflection  of light for the quintessence black hole. In a paper by Amore et.al.\cite{amore}, a new method was introduced to compute the deflection angle for a static spherically symmetric space-time. This method yields highly accurate analytical results. In the paper, they applied the technique to well known metrics. One of them was the metric in Weyl gravity \cite{pir1}\cite{pir2}. In Weyl gravity, the metric is given by,
\begin{equation}
ds^2 = -g(r) dt^2 + g(r)^{-1} dr^2 +  r^2 ( d \theta^2 + sin^2\theta d \phi^2)
\end{equation}
\begin{equation}
g(r)= 1 - \frac{ 2 \beta } { r}   + \gamma r - k r^2
\end{equation}
Even though the context in which the Weyl gravity and the quintessence black 
hole are formulated are different, one can see that for $k =0$ and $ \gamma = -c$, the Weyl gravity geometry and the quintessence black hole geometry is the same!. Therefore, the bending angle obtained by Amore et.al\cite{amore} for the Weyl gravity can be modified to obtain the angle for the quintessence black hole as,
$$
\alpha_{quintessence} = \frac{ 4 M}{ r_{o}} + 
\frac{ 4 M^2}{ r_{o}^2}\left( \frac{ 15 \pi}{16} -1 \right) $$

\begin{equation}
+ c ( r_o - M + \frac{ 3 \pi M}{2} ) + c^2 \frac{r_o^2}{2}
\end{equation}
Note that here, $r_o$ is the one given in eq.(57) for the quintessence black hole.

When $ c \rightarrow 0$, one obtain the  bending angle for the  Schwarzschild black hole, as,
\begin{equation}
\alpha_{Sch} = \frac{ 4 M}{ r_o^{Sch}} + 
\frac{ 4 M^2}{( r_o^{Sch})^ 2}\left( \frac{ 15 \pi}{16} -1 \right) 
\end{equation}
The bending angles, after substituting for $r_o$ for respective black holes can be plotted as a function of the impact parameter $D$. The graph is given in Fig.20.

\begin{center}
\scalebox{.9}{\includegraphics{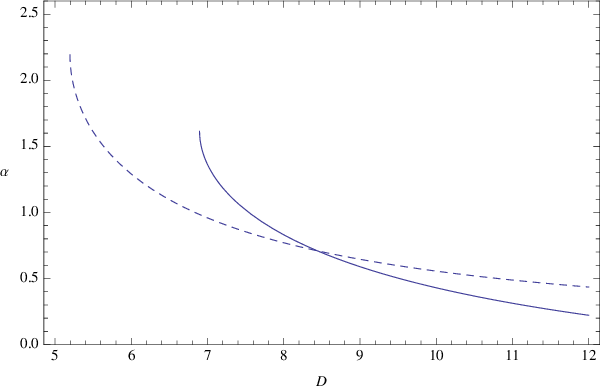}}\\
\vspace{0.2cm}
\end{center}
Figure 20. The graph shows the bending angle $\alpha$ for the quintessence black hole(dark curve)  and the Schwarzschild black hole(dashed curve)  as a function of the impact parameter $D$. Here, $M = 1$  and $c = 0.05$. $D_c = 6.8956$ for the quintessence black hole and $D_c = 5.19615$ for the Schwarzschild black hole.\\

From  Fig.20,  the photons with the same impact parameter bends more around the quintessence  black hole for small $D$ and less for large $D$. We like to mention here that Liu et.al \cite{liu} did studied the gravitational frequency shift and deflection of light for the black holes surrounded by the quintessence for various values of the state parameter $\omega_q$. There, the function $u$ was computed as a first order approximation where as here the exact solution is used.

\section{Conclusion}

We have studied the null geodesics of a black hole surrounded by dark energy. Here, we have chosen quintessence as the candidate for dark energy and the black hole solution studied was derived by Kiselev \cite{kiselev}. The equations for the geodesics were solved exactly for various values of energy and angular momentum of the photons. All possible motions are presented. The circular orbits are studied in detail and are shown to be unstable. We have also computed the Lyapunov exponent $\lambda$ which gives the instability time scale for the unstable geodesics. It is shown that $\lambda$ is smaller for the dark energy black hole in comparison with the $\lambda$ for the unstable circular orbits for the Schwarzschild black hole.

As an application of the photon motion studied here, we have studied the light deflection for this particular black hole and have done comparisons with the Schwarzschild black hole. The photons with the same impact parameter bends more around the quintessence  black hole for small $D$ and less for large $D$.

As an extension of this work, one could study the motion of particles around the charged version of the quintessence black hole.

\vspace{0.5cm}

{\bf Acknowledgment}: This work was supported in part by the Faculty  Summer Fellowship (2010)  of Northern Kentucky University.

\end{document}